# Accounting for objective lens autofluorescence in quantum emitter measurements

K. G. Scheuer, G. J. Hornig, T.R. Harrison, and R. G. DeCorby*

ECE Department, University of Alberta, 9211-116 St. NW, Edmonton, AB T6G 1H9, Canada
*Corresponding author: rdecorby@ualberta.ca

**Abstract**

The rise of interest in the study of quantum emitters has recently prompted many research groups to construct their own confocal epifluorescence microscopy/spectroscopy instruments. The low light levels typically involved in quantum emitter measurements makes it critically important to account for any potential sources of background fluorescence by components used within such setups. In this report, we show that emission originating from various microscope objectives can possess elusive sharp linewidths that could potentially be mistaken for quantum emission. Impurities present in many glasses overlap in wavelength with single photon emission from several candidate emitter systems of current interest. A careful consideration of this system noise could be critical to exploratory work of the optical properties of materials such as transition metal dichalcogenides and hexagonal boron nitride.

## Introduction

The recent increase of interest in quantum emitters in 2D materials such as transition metal dichalcogenides (TMDCs) [1] and hexagonal boron nitride (hBN) [2] has prompted many to construct homebuilt optical systems often targeting photoluminescence microscopy/spectroscopy as a particularly important measurement. Single photon emission (SPE) from such materials has a broad range of potential applications in quantum cryptography, communication, and computing [3]. However, because of the small-signal nature of many emitting materials, it is often a challenge to locate and correct for background fluorescence within a setup. The pieced-together nature of a homebuilt system adds an additional layer of complexity. In this short report, we show that microscope objectives, even those specifically designed for fluorescence applications, can be a significant source of background signal within many epifluorescence setups. The issue of "autofluorescence" within objectives has been widely studied in the context of fluorescence imaging [4–6], where it is typically treated as background noise that interferes with high contrast measurements. However, there is minimal discussion in the literature with respect to the spectral properties of this autofluorescence, which in the context of quantum emitters can be viewed as a ubiquitous and characteristic source of system noise. Moreover, when considering small-signal spectral measurements common in experiments involving SPEs, such background signals can approach the same order of magnitude as that from the sample. Thus, it is critically important to characterize and remove this background autofluorescence from any photoluminescence measurement of quantum emitters.

## Experimental Results

Figure 1 shows a comparison of fluorescence obtained from various microscope objectives (Table 1) using a 405 nm CW laser (Pangolin LDX-405NM-200MW) as an excitation source. Each spectrum is representative of a time-stable system autofluorescence that appears to depend on the objective lens used. Some of the spectra contain sharp lines while others exhibit only broad background features. The data presented is as measured (i.e., no averaging or post-processing) so that the amplitude of each spectrum can be directly compared. These measurements were taken using the homebuilt quasi-confocal photoluminescence setup shown in Fig. 2. This system was designed to measure defect-center emission from hBN, which spans a wide range of visible and near-infrared wavelengths (~ 500 nm – 850 nm) [2].

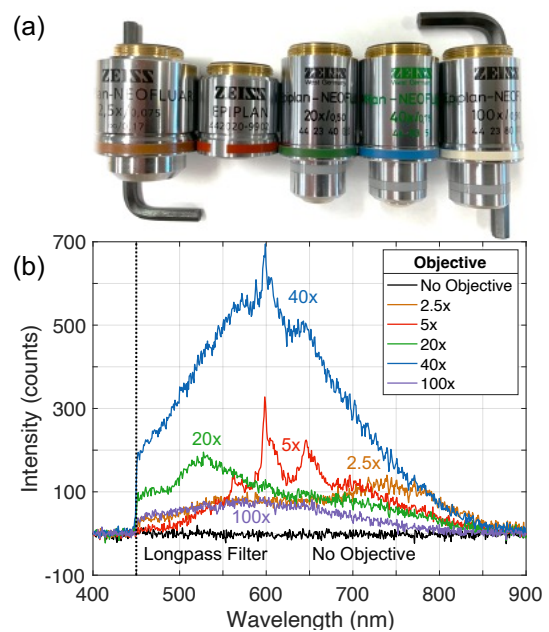

**Fig. 1. (a)** A photograph of the microscope objectives studied. **(b)** Representative photoluminescence signals for various objective lenses contained within a fluorescence microscopy/spectroscopy setup (Table 1) under 20 mW of 405 nm CW excitation. The sharp cut-off at 450 nm is due to the presence of a long pass filter in the collection path.

**Table 1. A list of the infinity-corrected Zeiss microscope objectives studied.**

| Magnification | Numerical Aperture | Objective[a] |
|---|---|---|
| 2.5x | 0.085 | Zeiss EC Plan-Neofluar |
| 5x | 0.13 | Zeiss Epiplan |
| 20x | 0.5 | Zeiss Epiplan-Neofluar |
| 40x | 0.75 | Zeiss Epiplan-Neofluar |
| 100x | 0.9 | Zeiss Epiplan-Neofluar |

[a]Although the objectives are not all from the same series, they are all listed as photoluminescence-grade by the manufacturer.

As with any fluorescence measurement, one primary goal is to efficiently deliver pump light to the desired region of the sample, while eliminating (as much as possible) its presence in the detection path. To that end, the excitation source was first sent through a neutral density filter (Newport FBS series) and an adjustable iris for attenuation and spatial filtering, then passed through a 450 nm cut-off short pass filter and a 468 nm cut-off short pass filter (Thorlabs FESH0450 and Semrock FF01-468, respectively). The filtered pump beam was reflected by a dichroic mirror (Thorlabs DMLP425R) towards the objective under test. In a typical measurement of a sample placed at the focus of the objective, some of the light emitted by the sample is collected back into the objective lens and then passed upwards through the dichroic mirror. To illustrate the role of microscope objective autofluorescence as a potential source of noise, we replaced the sample with an angled silver mirror. This mirror was located at a large distance (~ 10 cm) below the objective to minimize the possibility of scattered/reflected light coupling back into the system while also ensuring that it did not sit at the focal plane of any objective, so that potential luminescence from any surface debris was not collected. The collection side of the system (above the dichroic) was configured in the same way as it would be for typical fluorescence measurements. The collection path contained a 450 nm cut-off long pass filter (Thorlabs FELH0450) to further remove residual pump photons. The filtered light was delivered to a 90:10 beamsplitter (Thorlabs BS025); a CMOS camera (Thorlabs CS895MU) received 10% of the split beam, while the remaining 90% was sent towards a 50 μm-core pickup fiber (Thorlabs FG050LGA) mounted to a 34.74 mm focal length fiber collimator (Thorlabs F810FC-543). The output of the fiber was coupled directly into a USB spectrometer (Ocean Optics QEPro) configured with a 25 μm slit.

To investigate the variation of autofluorescence signal with pump power, we collected spectra for pump powers in the 2 mW ~ 50 mW range using the 5x objective lens. The 5x lens was selected because of its particularly sharp spectral features. Figure 3(a) shows the evolution of fluorescence spectra for varying pump laser power. The relationship between pump power and the fluorescence signal power is given in the inset of Fig. 3(a), where the reported fluorescence signal power is integrated over all wavelengths. A linear fit to this data revealed a slope of 163 ± 1 counts s$^{-1}$ mW$^{-1}$ with $R^2$ = 0.9985. While the collected power is system-dependent, the linear relationship between the fluorescence and illumination ($I_{flu} \propto I_i$) is evidence of emission below saturation [7].

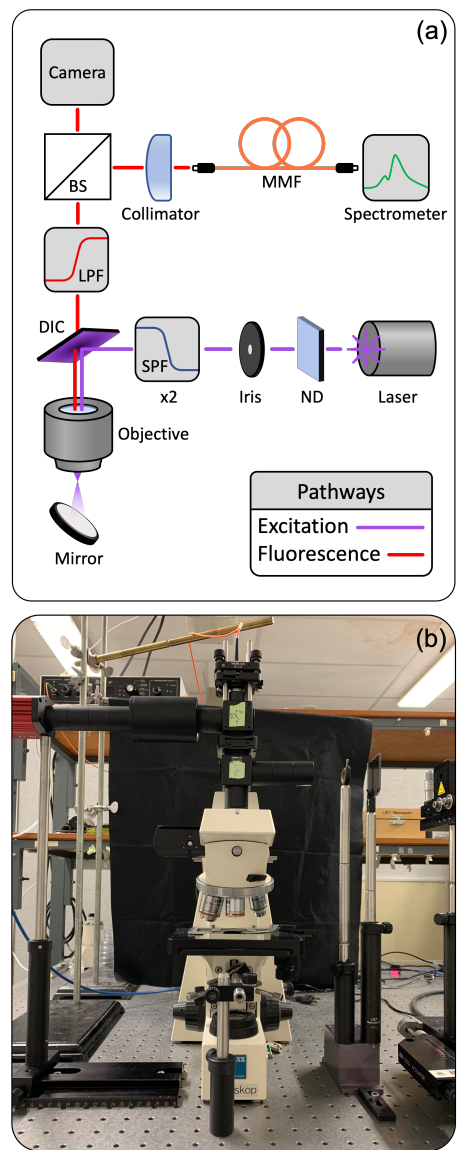

**Fig. 2.** Experimental setup. **(a)** A schematic of the experimental setup used to perform photoluminescence measurements with no sample in place. An angled silver mirror was placed far below the focal plane of the objective lens to minimize collection of back-scattered pump photons. (Legend: DIC – dichroic mirror, ND – neutral density filter, BS – beamsplitter, MMF – multimode fiber, SPF/LPF – short pass/long pass filter). **(b)** A photograph of the experimental setup.

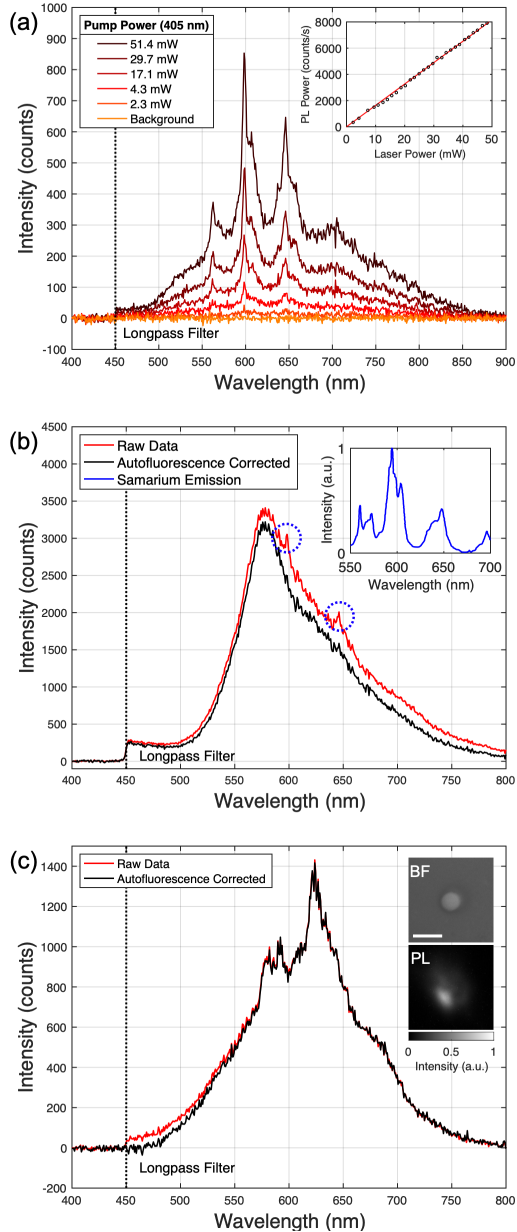

**Fig. 3. (a)** Effect of pump power on photoluminescence spectrum for the 5x objective from Table 1. The laser pump power incident on the objective was adjusted as the resulting photoluminescence signal was collected at each step. The inset shows pump laser power versus photoluminescence power for the same objective (black circles). Photoluminescence power was obtained by integrating over the detection range for each spectrum. An approximately linear relationship (red line) was observed, with a slope of $163 \pm 1$ counts $s^{-1}$ $mW^{-1}$. **(b)** Photoluminescence obtained from a very large ensemble of hBN emitters with (black) and without (red) autofluorescence correction applied captured using the 5x objective (Table I) with ~ 20 mW of pump power. Blue circles contain peaks likely corresponding to objective lens autofluorescence. The inset shows an artistic representation of photoluminescence (blue) obtained from samarium defects in natural fluorite excited with a 400 nm pump [8]. **(c)** Photoluminescence obtained from an ensemble of hBN emitters on a 3.6 μm patterned $SiO_2$ pillar (described in previous work [9]) with (black) and without (red) autofluorescence correction applied captured using the 100x objective (Table 1) with ~ 3 mW of pump power. Bright field (BF) and photoluminescence (PL) images of the pumped region are present as insets with a 5 μm scale bar.

## Discussion

As is clear from Fig. 1, even with no sample of any kind present, spectra intrinsic to the experimental setup are observed and appear to be unique to each objective lens. In all cases, the emission extends from < 450 nm to ~ 850 nm, and in some cases contains sharp, distinct spectral features. The most likely sources of this apparent emissive property of the objective lenses, and specifically the sharp emission lines centered at ~ 600 nm produced by the 5x and 40x objectives, are defects and impurities within the objective glass itself. Fluorite glass, among many others, has long been known to contain luminescing defects [8,10–12]. Most of the objective lenses used in this study contain fluorite, though the exact compositional breakdown is not provided by the manufacturer. It is well known that fluorite and other host materials contain dopants/impurities such as $Sm^{3+}$, $Dy^{3+}$ and $Eu^{3+}$ which have atomic transitions near 600 nm [8,13–16]. For example, L-band excitation (~ 400 nm) of fluorite and crown glass containing $Sm^{3+}$ produce emission spectra [8,16] that almost exactly match that observed from the 5x and 40x objectives lenses in this study [8]. Sharp emission lines in this range have significant potential to interfere with those of SPEs hosted by hBN and other materials, a fact which has not been widely discussed in the context of quantum emitters. To add to the confusion, one of the hallmarks of some types of single photon emission in hBN is the energy spacing between a narrow zero phonon line and a broader phonon sideband (ΔE ~ 160 meV or ~ 50 nm) [17]. Some of the data from the 5x and 40x objectives could easily be mistaken for such features. For example, consider the sharp feature near 650 nm and broad feature centered about 700 nm in Fig. 3(a). To demonstrate this point, we have included an emission spectrum from a continuous hBN film (Graphene Supermarkets CVD-2X1-BN-ML) that was transferred to a pre-patterned $SiO_2$ substrate using a polymer-assisted transfer process [9]. A large ensemble of emitters measured using the 5x objective is shown in Fig. 3(b). The raw, uncorrected spectrum (red) contains the same interference signals (circled blue) present in Fig. 3(a) that we attribute to the objective lens. Performing a background subtraction with no sample (black) using the spectra in Fig. 3(a) almost entirely removes these features while leaving other spectral content undisturbed. If not accounted for, such autofluorescence signals have potential to obscure the identification of single photon emitters. However, with proper alignment and high collection efficiency this background autofluorescence can be reduced to a relatively innocuous level, for example as shown in Fig. 3(c), where an ensemble of emitters was measured using the 100x objective.

We would also like to emphasize that such interference signals could be particularly problematic in epifluorescence systems that use fast steering mirrors to direct the pump laser, since the position and angle of the beam on the objective would vary, resulting in autofluorescence that appears to correlate with position on a sample. High sample reflectivity (at the pump wavelength) could also greatly increase the autofluorescence signal in some cases.

The fact that each objective has a unique emissive characteristic is likely owed to a combination of effects involving their differing physical construction (i.e., size, material composition) and optical properties (i.e., working distance, numerical aperture) further complicating the process of identifying, tracking, and calibrating for this autofluorescence. In addition, other optical components within the epifluorescence system such as the dichroic mirror, beamsplitter, and fiber pinhole have potential to further colour the autofluorescence signal. Photoluminescence measurements of the objective lenses in isolation are presented in Supplementary Information to demonstrate this point.

We are confident in stating that laser non-idealities are not the source of the signals we observed here. Multi-stage input filtering was used to suppress spectral content above ~ 450 nm wavelength prior to the beam entering the objective turret. The effectiveness of this filtering was verified by addition of extra short-pass excitation filters, confirming that this had no impact on the measured spectra above. Furthermore, we found that replacing the objective under test with a silver mirror eliminated the photoluminescence signal, as shown in Fig. 1(b), further confirming that the photoluminescence spectra shown are attributable to emission by the objective lenses.

## Conclusions

In summary, we have demonstrated that background fluorescence from microscope objective lenses can be a potentially problematic source of interference in the context of quantum emitters. Certain materials like samarium contained within objective lens glass might produce sharp emission lines which can be mistaken for defect-based emission. The important message is that researchers need to be mindful of all materials contained within the beam path of their photoluminescence systems when measuring small signals. Performing a system characterization as shown in Fig. 2(a) may yield concerning but nonetheless necessary results.

## Acknowledgments

This work was supported by the Natural Sciences and Engineering Research Council of Canada (CREATE 495446-17); Alberta Innovates (Graduate Student Scholarship); Alberta EDT Major Innovation Fund (Quantum Technologies); and Carcross/Tagish First Nations. We thank Wilson Analytical Services Inc. for their generous sharing of equipment.

# Accounting for objective lens autofluorescence in quantum emitter measurements: supplementary information

K. G. Scheuer, G. J. Hornig, T.R. Harrison, and R. G. DeCorby*

ECE Department, University of Alberta, 9211-116 St. NW, Edmonton, AB T6G 1H9, Canada
*Corresponding author: rdecorby@ualberta.ca

immediately after the exit pupil of the objective to filter pump light. The photoluminescence signal was collected using a standard small-format collimator (Thorlabs) and passed into the Ocean Optics QEPro spectrometer with a 600 μm core fiber (Thorlabs). Spectra collected using a pump power of ~ 100 mW and integration time of 1 second are presented in Fig. S1 along with a schematic of the measurement setup. The data is again presented as measured.

Removing extraneous optical elements from the measurement such as the dichroic mirror, beamsplitter, and fiber pinhole provided a more direct measurement of the objective lens luminescence and shows that spectral features centered near 600 nm and 650 nm are generally present across all objective lenses. Additionally, no detectable signal is present without an objective in place, as shown in Fig. S1(b).

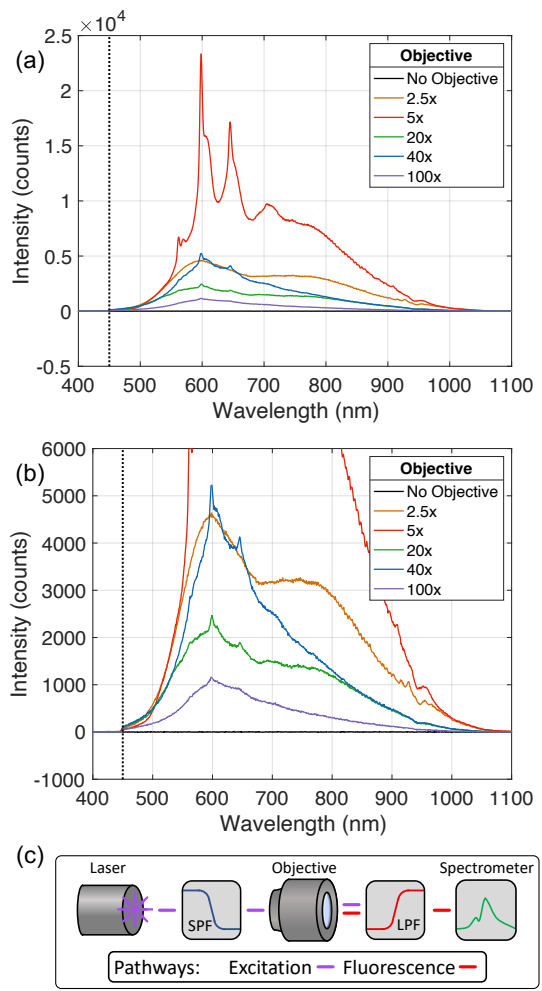

**Fig. S1.** Direct measurements of photoluminescence from the microscope objectives in Table 1. **(a)** Measurements for each objective subject to ~ 100 mW of power captured using an integration time of 1 second. **(b)** A zoomed version of (a) showing subtle differences between the 2.5x, 20x, 40x, and 100x objectives as well as no appreciable signal in the case where no objective was present. **(c)** A simplified schematic of the isolated experimental setup. (Legend: SPF/LPF – short pass/long pass filter).

To further demonstrate that the persistent photoluminescence signal was attributable to the objective lenses a secondary experiment was performed that involved isolating the lenses from other components within the epifluorescence system. In this case, the excitation laser was filtered using the same short pass filter set described in the main text and passed directly into the collection side of the objective under test. Two long pass filters (both Thorlabs FELH0450) were placed